# Quantum Humeanism, or: physicalism without properties

Michael Esfeld
University of Lausanne, Department of Philosophy
Michael-Andreas.Esfeld@unil.ch

**Abstract**

In recent literature, it has become clear that quantum physics does not refute Humeanism: Lewis's thesis of Humean supervenience can be literally true even in the light of quantum entanglement. This point has so far been made with respect to Bohm's quantum theory. Against this background, this paper seeks to achieve the following four results: (1) to generalize the option of quantum Humeanism from Bohmian mechanics to primitive ontology theories in general; (2) to show that this option applies also to classical mechanics; (3) to establish that it requires a commitment to matter as primitive stuff, but no commitment to natural properties (physicalism without properties); (4) to point out that by removing the commitment to properties, the stock metaphysical objections against Humeanism from quidditism and humility no longer apply. In that way, quantum physics strengthens Humeanism instead of refuting it.

## 1. Introduction

For a long time, it was thought that quantum entanglement refutes Humeanism, in particular David Lewis's thesis of Humean supervenience (e.g. Lewis 1986: introduction), because the entangled wave-functions of quantum systems are not compatible with an ontology that admits only local matters of particular fact. That incompatibility was supposed to follow from Bell's theorem (Bell 1964, reprinted in Bell 1987: ch. 2) and the subsequent experiments (e.g. Aspect, Dalibard and Roger 1982).

Thus, Teller (1986) claimed that quantum entanglement requires recognizing relations that do not supervene on the local matters of particular fact and thereby rules out local physicalism. Teller's non-supervenient relations were later spelled out in terms of ontic structural realism, and Teller's view of these relations being instantiated by individuals was abandoned: the entangled wave-functions of quantum systems were seen as committing us to an ontology of concrete physical structures in the sense of relations that (a) do not supervene on intrinsic properties of their relata and that, moreover, (b) do not even require that their relata have an intrinsic identity at all (e.g. Ladyman 1998, French and Ladyman 2003, Esfeld 2004). Furthermore, these structures were considered to be modal, establishing connections in nature that are in any case more than mere contingent regularities, if not outright necessary connections. Consequently, quantum entanglement was considered as refuting not only the idea of an ontology that admits only local matters of particular fact, but also the core tenet of any form of Humeanism, namely to eschew a commitment to objective modality (see notably

Maudlin 2007: ch. 2, in particular pp. 51-64; see Ladyman and Ross 2007: chs. 2-5, Esfeld 2009, French 2014: chs. 9-11 on modal structures). The commitment to only local matters of particular fact follows from the intention to do without objective modality: maintaining that nature is nothing more than a mosaic of local matters of particular fact is the most straightforward way to avoid any sort of a commitment to modal connections in nature.

Philosophers with a favourable attitude towards Humeanism reacted to this situation by trying to adapt Humeanism so that quantum entanglement is taken into account. The most important suggestion in this respect is to admit irreducible relations over and above the spatio-temporal relations to the ontological ground floor of Humeanism (Darby 2012) and to envisage developing a Humean version of ontic structural realism on the basis of including such relations (Lyre 2010). However, recognizing irreducible relations of quantum entanglement considerably restricts free combinatorialism and arguably implies a commitment to some sort of objective modality, since these relations tie the temporal development of – in the last resort all – quantum systems together, whatever their spatial distance may be. Thus, considering the experiment of Einstein, Podolsky and Rosen (EPR) (1935) in the version of Bohm (1951: 611-22), if in one wing of the experiment the measured quantum system behaves in such a way that the measurement outcome is spin up, then it is necessarily so that in the other wing of the experiment the measured quantum system behaves in such a way that the measurement outcome is spin down. Consequently, if there are relations of quantum entanglement in the supervenience base, these relations pose a constraint on what can and what cannot happen elsewhere in space-time.

Furthermore, in order to rescue Humeanism, one may suggest that the very high-dimensional configuration space of the universe is the realm of physical reality (Loewer 1996) (if there are $N$ particles in three-dimensional space, the dimension of the corresponding configuration space is $3N$). Hence, in this case, the physical world is not situated in a three-dimensional space or a four-dimensional space-time, but plays itself out in a very high-dimensional space. If one makes this move, there is no problem for Humeanism, since the quantum mechanical wave-function can be considered as a field in configuration space, assigning values to the points of that space that can be regarded as intrinsic properties occurring at the points of that space. Moreover, the development of the wave-function is local in that space, as long as it is given by a linear dynamical equation (such as the Schrödinger equation) (see Albert 1996 and 2013 as well as Ney 2010: section 3.3).

However, this adaptation implies taking what is usually introduced as a mathematical space *representing* the physical reality – with each point of $3N$ dimensional configuration space representing a possible configuration of $N$ particles in three-dimensional space – to be *itself* the physical reality. Note that this suggestion is different from the proposals put forward in the context of the search for a quantum theory of gravity according to which physical space may have more than four dimensions (as e.g. in string theory): these proposals do not call into question the contrast between a configuration space as a mathematical space employed to represent the physical reality and physical space, even if physical space should turn out to be different from what is currently presupposed in quantum field theory, or general relativity theory. Configuration space realism therefore entails that one has to give up a central tenet not only of common sense realism, but also of all working science, namely the one of a distinction between a mathematical space that is employed to represent physical reality and the space in which physical reality is situated. One can argue that this tenet should be given up only as a last resort (Monton 2006).

Fortunately for the Humean, in recent years, it has become clear that no such adaptation is necessary. Humeanism is not refuted by quantum physics. More precisely, Lewis's thesis of Humean supervenience can be literally true even in the light of the empirical evidence for quantum entanglement. The background that enables Humeanism to stand firm is the development of what is known as primitive ontology theories of quantum physics. The term "primitive ontology" goes back to Dürr, Goldstein and Zanghì (2013: ch. 2, end of section 2, originally published 1992). Attention has focused in recent years on the structure of these theories, following notably the paper by Allori et al. (2008): the primitive ontology consists in the distribution of matter in three-dimensional space or four-dimensional space-time; that distribution is the referent of the formalism of quantum physics. Furthermore, a law is admitted as that what fixes (in a probabilistic or a deterministic manner) the temporal development of the distribution of matter in physical space, given an initial configuration of matter. That's all. In particular, the quantum mechanical wave-function is part and parcel of the law instead of being a physical entity on a par with the primitive ontology.

The primitive ontology is in any case constituted by local matters of particular fact – "local beables" to use Bell's famous neologism (Bell 1987: ch. 7) ("beable" standing for what there is by contrast to "observable", i.e. what can be observed). The only move that the Humean has to make then is this: instead of admitting the law as an entity that exists in addition to and independently of the primitive ontology, governing or guiding the temporal development of the primitive ontology, the Humean has to regard the law as supervening on the distribution of matter throughout the whole of space-time, that is, the entire mosaic of "local beables" or local matters of particular fact. This move has been made with respect to Bohm's quantum theory in recent literature (Callender unpublished, Esfeld et al. 2013: section 3, Miller 2013).

Against the background of this state of the art in the recent literature, the aim of the present paper is to achieve the following four results:

- to generalize the option of quantum Humeanism from Bohm's quantum theory to primitive ontology theories of quantum physics in general (section 2);
- to show that this option is applicable not only to quantum physics, but also to classical mechanics (section 3);
- to establish that it requires a commitment to matter as primitive stuff, but no commitment to natural properties (physicalism without properties) (end of section 2);
- to point out that by removing the commitment to natural properties, the stock metaphysical objections against Humeanism from quidditism and humility no longer apply (section 4).

In a nutshell, far from refuting Humeanism, quantum physics strengthens Humeanism as a stance in the metaphysics of science. The only amendment that the Humean has to make is to endorse a commitment to primitive stuff instead of intrinsic properties occurring at space-time points.

2.    *Humeanism and the primitive ontology of quantum physics*

There are three elaborate primitive ontology theories of quantum mechanics. The de Broglie-Bohm theory, going back to de Broglie (1928) and Bohm (1952) is the oldest of them. Its dominant contemporary version is known as Bohmian mechanics (Dürr, Goldstein and Zanghì 2013). This theory endorses particles as the primitive ontology, maintaining that there is at any time one configuration of particles localized in three-dimensional space, with the particles moving on continuous trajectories in space. Bohmian mechanics therefore needs two laws: the guiding equation fixing the temporal development of the position of the particles, and the Schrödinger equation determining the temporal development of the universal wave-function, that is, the wave-function of all the particles in the universe. These two laws are linked in this way: the role of the universal wave-function, developing according to the Schrödinger equation, is to determine the velocity of each particle at any time $t$ given the position of all the particles at $t$.

Furthermore, there are two primitive ontology theories using the dynamics proposed by Ghirardi, Rimini and Weber (GRW) (1986), which seeks to include the textbooks' postulate of the collapse of the wave-function upon measurement into a modified Schrödinger equation. Employing this dynamics, Ghirardi, Grassi and Benatti (1995) develop an ontology of a continuous matter density distribution in physical space: the wave-function in configuration space and its temporal development as described by the GRW equation represent at any time the density of matter in physical space, such that there is at any time a certain matter density instantiated at the points of physical space. The spontaneous localization of the wave-function in configuration space (its collapse) represents a spontaneous contraction of the matter density in physical space, thus accounting for measurement outcomes and well localized macrophysical objects in general (see Monton 2004 for details).

The other theory goes back to Bell (1987: ch. 22): whenever there is a spontaneous localization of the wave-function in configuration space, that development of the wave-function in configuration space represents an event occurring at a point in physical space. These point-events are today known as flashes; the term "flash" was introduced by Tumulka (2006: p. 826). According to the GRW flash theory, the flashes are all there is in space-time. Consequently, the temporal development of the wave-function in configuration space does not represent the distribution of matter in physical space. It represents the objective probabilities for the occurrence of further flashes, given an initial configuration of flashes. There thus is no continuous distribution of matter in physical space, namely no trajectories or worldlines of particles, and no field – such as a matter density field – either. There only is a sparse distribution of single events in space-time.

These theories hence put forward different proposals about the nature of matter, which cover the main metaphysical conceptions of matter – particles, gunk, single events. That notwithstanding, their structure is the same: they consist in a proposal for a primitive ontology of matter distributed in physical space and a law for its temporal development (see Allori et al. 2008). The proponents of these theories tend to regard the universal wave-function as nomological – that is, as part of the law – instead of considering it as a physical entity that exists in physical space in addition to the primitive ontology (see notably Dürr, Goldstein and Zanghì 2013: chs. 11.5 and 12). The reason for adopting a nomological stance with respect to the universal wave-function is that it cannot be conceived as a wave or field in physical space: it does not have values at the points of physical space. If it is a wave or a field, it can be a wave or field only in configuration space, that is, the very high-dimensional mathematical space each point of which corresponds to a possible configuration of matter in physical space.

Nonetheless, the proponents of these theories are inclined to consider the law – and in particular the universal wave-function – as an entity that exists in addition to the primitive ontology, being some sort of a "non-local beable" that there is over and above the "local beables" which constitute the primitive ontology (see again Dürr, Goldstein and Zanghì 2013: chs. 11.5 and 12). The main reason for doing so is that the primitive ontology is not sufficient to determine the universal wave-function: there can be two or more identical initial universal configurations of particles, matter density or flashes and yet different wave-functions of these configurations, leading to different temporal developments of them.

However, the Humean does not claim that the laws of nature supervene on the local matters of particular fact *at a given time*. The Humean claim is that the laws of nature supervene on the entire distribution of the local matters of particular fact throughout the *whole* of space-time. In other words, the laws of nature are not fixed when there is the initial configuration of local matters of particular fact; they are determined only at the end of the world so to speak (see e.g. Beebee and Mele 2002: 201-5). Accordingly, the Humean claim is that the universal wave-function is determined only by the distribution of the particle positions, the matter density or the flashes throughout the whole of space and time. Whereas it is not an option to hold that the universal wave-function supervenes on the configuration of the local matters of particular fact at any given time, it is an option to maintain that it supervenes on the local matters of particular fact throughout the whole of space and time. If the entire distribution of the local matters of particular fact were still to leave room for different universal wave-functions, that difference would not make any empirical difference and could therefore be regarded by the Humean as a mathematical surplus structure.

In making this move, the Humean has to give up the idea that the universal wave-function guides or pilots the temporal development of the configuration of matter in space. The Humean can invoke a good reason for abandoning that idea: since the universal wave-function cannot be a wave or field that exists in physical space, but only a wave or field in configuration space, it is in any case unintelligible how an entity in configuration space could guide or pilot the temporal development of entities in physical space. There is a connection of representation, but certainly not a causal connection between configuration space and physical space, with an entity existing in the former space piloting entities existing in the latter space.

To put it differently, the Humean can with good reason argue that the idea of the universal wave-function guiding or piloting the temporal development of matter in physical space is not part of the physics of, say, Bohmian mechanics, but reveals a metaphysical prejudice in favour of an anti-Humean conception of laws of nature being something that governs the behaviour of what there is in the physical world (cf. Beebee 2000). For the Humean, the universal wave-function and the dynamical law that appear in a physical theory such as Bohmian mechanics or the GRW theory are part of the best system, that is, the system that achieves the best balance between being simple and being informative in capturing what there is in the physical world. That system – and everything that belongs to it – supervenes on the entire distribution of the local matters of particular fact throughout the whole of space-time.

It is hence a coherent position to maintain that the primitive ontology is the *full* ontology, in the sense that everything else – including the universal wave-function and the law in which it figures – supervenes on the entire distribution of the "local beables" in the whole of space-time. Bell himself recognized this position as a coherent stance in the paper in which he introduced the notion of "local beables" (1975):

> One of the apparent non-localities of quantum mechanics is the instantaneous, over all space, 'collapse of the wave function' on 'measurement'. But this does not bother us if we do not grant beable status to the wave function. We can regard it simply as a convenient but inessential mathematical device for formulating correlations between experimental procedures and experimental results, i.e., between one set of beables and another. (Quoted from Bell 1987: 53)

Bell makes two important points in this quotation: (1) It is not mandatory to grant beable status to the wave-function. If one admits "local beables", one has an ontology of the physical world. Not granting beable status to the wave-function does, however, not commit one to an instrumentalist attitude to the wave-function, as Bell suggests here. Humeanism is distinct from instrumentalism (Miller 2013: section 5 stresses this point). The Humean only has to maintain that the primitive ontology is the full ontology, with everything else supervening on it. That is why Humeanism is also not touched by recent claims about experimental evidence in favour of the reality of the wave-function (Pusey, Barrett and Rudolph 2012; Colbeck and Renner 2012): these claims only seek to rule out the view that the wave-function represents nothing but the information about probabilities for measurement outcomes that is available for an observer. However, on Humeanism, the universal wave-function is not tied to an observer: it supervenes on the entire distribution of the local matters of particular fact in space-time. Since any experimental evidence consists in "local beables", in making that move, the Humean is in the position to accommodate whatever experimental evidence there may be.

(2) Given that it is the wave-function which is entangled and which correlates "local beables" whatever their spatial or spatio-temporal distance is, if one does not grant beable status to the wave-function, there is no reason to admit non-supervenient relations of entanglement (or of dependence or of influence) among the "local beables" over and above their occurrence at space-time points. In being entangled, the wave-function establishes such correlations, but these are no addition to what there is over and above the occurrence of the "local beables" at space-time points, since the universal wave-function and its temporal development supervene on the entire mosaic of these "local beables".

If one does not grant beable status to the wave-function, there is nothing that determines the temporal development of the initial configuration of the "local beables". The particle positions simply happen to develop in such a way that there are, as far as Bohmian quantum mechanics is concerned, continuous particle trajectories; the matter density values just happen to develop in such a way that the matter density takes a certain shape making true the GRW law, and the flashes just happen to occur in such a manner that they make true a law of the GRW type. There is nothing that drives, guides or forces them to do so. This is, of course, an instance of the general Humean attitude towards laws and objective modality. One may have reservations about that attitude. But there is nothing in quantum physics that obliges one to abandon it. In brief, it is "anti-Humeanism in, anti-Humeanism out", or "Humeanism in, Humeanism out". If one takes for granted that the wave-function is some sort of a real entity or "non-local beable" in addition to the primitive ontology, then quantum physics comes out anti-Humean. If, by contrast, one bases oneself on the empiricist idea that the primitive ontology is the full ontology, then one obtains a Humean ontology of quantum physics.

The primitive ontology theories are only concerned with the position of matter in space and its development in time. The theorems of Gleason (1957) and Kochen and Specker (1967) among others show that it is not possible to regard the quantum mechanical operators or observables as describing properties that the objects in nature possess, since one cannot attribute values to these observables independently of measurement contexts. The observables are not properties of anything. They are ways in which the quantum objects behave in measurement contexts. The primitive ontology theories account for the observables in terms of how the position of the objects in physical space develops in such contexts. Consider spin: these theories explain the outcomes of spin measurements in terms of the temporal development of the position of objects in physical space as described by the wave-function. This has been done in detail for Bohmian mechanics (see Bell 1987: ch. 4 and Norsen 2013). There is no doubt that the same treatment is available for the GRW matter density theory and the GRW flash theory (cf. Tumulka 2006, 2009).

Consequently, their position in space is the only property of the physical objects in the sense that to it alone a value is attributed. All measurement outcomes supervene on that value. The velocity of the Bohmian particles is not anything in addition to their position, it simply is the temporal development of their position as given by the guiding equation. For the Humean, the guiding equation, including the universal wave-function that figures in it, does not represent a property of anything over and above position either; it is a law that supervenes on the particles' positions throughout the whole of space-time.

However, attributing a value that indicates where the particles are in space does not imply that position is a property that is instantiated by something. The Bohmian particles are primitive stuff: a particle being at a space-time point simply signifies that the point is occupied instead of empty. Accordingly, the Bohmian particle configuration in space at any given time consists in certain points of space being occupied at that time, whereas other points are empty. The only difference between Bohmian particles and GRW flashes is that on Bohmian mechanics, points of space are occupied in such a way that the occupied points form continuous lines in time (worldlines). According to the GRW flash theory, by contrast, points are occupied in such a way that there are gaps in space as well as in time. Furthermore, the only difference between Bohmian particles and GRW flashes on the one hand and the GRW matter density on the other is that according to the latter, matter is distributed continuously throughout space in time – there are no empty points of space –, with there being more matter in some regions of space than in others. But the matter density simply is the density of stuff (gunk) (cf. Allori et al. 2013: 9-10).

In other words, the only variation that Bohmian mechanics and the GRW flash theory admit is the one of points of space being occupied or empty, with there being a change in time in which points are empty and which ones are occupied; the dynamical law, including the universal wave-function, is the description of that change which achieves the best balance between being simple and being informative. The only variation that the GRW matter density theory admits is the one of there being more stuff in some regions of space than in others, with there being a change in time in which regions of space there is more and in which ones there is less stuff; the dynamical law, including the universal wave-function, is the description of that change which achieves the best balance between being simple and being informative.

Hence, these theories are primitive ontology theories in yet another sense: they are committed to matter being primitive stuff distributed in space, and a law for the temporal development of that distribution. In brief, the primitive ontology also is primitive in the sense that it consists in primitive stuff. The stuff is primitive, because it does not have any properties. It simply is distributed in space, with its distribution varying in time. More precisely, the stuff does not have any physical or natural properties, given a sparse conception of properties that does not take predicates as a guideline for properties. Of course, one may attribute to the stuff "properties" such as the ones of being stuff or being matter and of being self-identical and the like. However, ascribing such "properties" to the stuff does not cut any ontological ice: it does not add anything to saying that there is primitive stuff distributed in space. For instance, endorsing such properties does not yield properties in the sense of ways of being of the stuff (cf. the ontology of sparse properties as ways of being of substances advocated by Heil 2012: chs. 4-5).

Instead of attributing properties to matter, one could contemplate conceiving these primitive ontology theories in the framework of super-substantivalism, according to which space-time is the only substance and matter a property of space-time: space-time points have the property of being occupied or being empty. However, it is doubtful how being occupied or being empty could be a *bona fide* property of space-time points, with the metrical properties as treated in general relativity theory setting the paradigm for what a *bona fide* property of space-time points is. Saying that space-time points have the property of being occupied or being empty amounts to what Sklar (1974: 166, 222-3) calls a linguistic trick, instead of vindicating super-substantivalism as a serious ontological position in the framework of primitive ontology theories of quantum physics. In brief, being occupied or being empty does not look like a property of anything, at least not on a sparse conception of properties. Being occupied signifies that there is something at the space-time point in question (namely stuff), whereas being empty signifies that there is nothing at the point in question.

Consequently, applying Humeanism to the primitive ontology theories of quantum physics after all entails a modification of Lewis's Humean ontology: the mosaic of local matters of particular fact does not consist in "local qualities: perfectly natural intrinsic properties which need nothing bigger than a point at which to be instantiated" (Lewis 1986: x). There are no such qualities or intrinsic properties in quantum physics. Quantum entanglement rules out that such properties could do any work as far as the features that are specific for quantum physics are concerned. Nonetheless, quantum entanglement notwithstanding, Lewis's thesis of Humean supervenience can be literally true in quantum physics, as the primitive ontology theories show. The only adaptation that is necessary to obtain this result is that the mosaic of local matters of particular fact is not constituted by local qualities occurring at space-time points, but by space-time points being occupied by primitive stuff or being empty.

This adaptation does not change anything as far as Humean supervenience is concerned, since what does all the work for the Humean supervenience thesis is that there is – only – a mosaic of local matters of particular fact. Conceiving that mosaic in terms of qualitative, intrinsic properties does not do any work, because the qualitative nature of these properties is in any case a pure quality, known as a quiddity. Consequently, we cannot have any epistemic access to such pure qualities. This consequence is known as humility. Lewis (2009) endorses both quidditism and humility, but these commitments put a rather heavy burden on Humeanism, in particular given that Humeanism sees itself as a metaphysics that is close to science and empiricism, avoiding any sort of occult metaphysics (see notably Black 2000 against quidditism). The mosaic of local matters of particular fact is needed as the supervenience basis for everything: it makes true laws and causal statements as well as all the functional descriptions of what there is in the world, captured in the Ramsey sentence of the world. That sentence quantifies over the whole distribution of local matters of particular fact in space-time, with that distribution realizing all the functions that there are in the world. But a mosaic of local matters of particular fact consisting in certain space-time points being occupied whereas others are empty can do so in the same way as a mosaic of local matters of particular fact consisting in pure qualities occurring at space-time points: it can make true laws – and thereby causal statements – as shown above. Furthermore, whatever functions there are in the world can be realized by configurations of primitive stuff that implement a variation consisting in certain space-time points being occupied, whereas others are empty. If *a priori* physicalism is true, then if one had complete knowledge of which space-time points are occupied throughout the whole of space-time, one could deduce from that knowledge *a priori* all the true statements about what there is in the world. In a nutshell, thus, applying Humeanism to the primitive ontology theories of quantum physics amounts to physicalism without properties.

*3.    Humeanism and the primitive ontology of classical mechanics*

The term "primitive ontology" has been introduced in the context of quantum physics. Nonetheless, primitive ontology theories apply wherever physics applies. By the same token, the combination of Humeanism with primitive ontology theories, resulting in physicalism without properties, is a live option not only for quantum physics, but for physics in general. Consider classical mechanics: again, atoms in the sense of Newtonian particles are primitive stuff. A particle being at a point in space at a certain time signifies that the point is occupied instead of empty. All the variation that there is in time is a change in which points of space are occupied and which ones are empty as time passes. Again, that change is such that there are continuous lines of occupation (worldlines interpreted as particle trajectories).

In the laws that describe that change figure variables standing for mass and charge, which are commonly regarded as intrinsic properties of the particles. Indeed, mass and charge are the paradigm examples of "local qualities: perfectly natural intrinsic properties which need nothing bigger than a point at which to be instantiated" (Lewis 1986: x). Given the distribution of mass and charge in space (plus certain constants such as the gravitational constant), the trajectories of the particles are fixed by the laws of classical mechanics. No further ontological commitment to forces or fields is necessary. However, mass and charge are admitted in classical mechanics only because they perform a certain function as described by the laws of gravitation and electromagnetism, namely to accelerate the particles in a certain manner. For a Humean, it is contingent that mass and charge exercise that functional role in the actual world. That they do so is by no means essential to them, but supervenes on the mosaic of local matters of particular fact as a whole, that is, the distribution of mass and charge qua local qualities throughout the whole of space-time. However, the qualitative nature of mass and charge does not do any job in Humeanism: it is a pure quality (quiddity) to which we moreover have no epistemic access (humility).

Hence, assuming that the mosaic of local matters of particular fact consists in the instantiation of qualitative, intrinsic properties at space-time points does not do any job in Humeanism applied to classical mechanics either. It is sufficient for the purpose of having a Humean supervenience basis to maintain that the mosaic of local matters of particular fact consists in some points of space-time being occupied, whereas others are empty. In short, a variation in the sense of some points of space being occupied whereas others are empty (primitive stuff) with a change in time in which points are occupied and which ones are empty is sufficient to constitute the mosaic of local matters of particular fact that serves as a supervenience basis for everything on Humeanism – in a world in which classical mechanics is the fundamental physical theory as well as in a quantum world.

Of course, if one assumes that there is something in nature that determines the temporal development of the primitive stuff, then one is committed in classical mechanics to particles being not only primitive stuff, but instantiating properties such as mass and charge, with it then being essential for these properties to accelerate the particles as described by the laws of gravitation and electromagnetism. But if one sides with the Humean in maintaining that there is nothing in nature which carries out such a determination, there is no need for a commitment to these properties being instantiated in nature in addition to facts about which space-time points are occupied and which ones are empty (see Hall unpublished: §5.2). There is a mass variable and a charge variable in the laws of gravitation and electromagnetism. However, these laws and everything that figures in them supervene on the mosaic of local matters of particular fact throughout the whole of space-time. They do not refer to anything that exists in nature over and above that mosaic, which can simply consist in some space-time points being occupied by primitive stuff and others being empty. In other words, given the laws, one can attribute properties like mass and charge to the primitive stuff – that is, the particles in classical mechanics. But the particles do not have these properties *per se*, as something essential or intrinsic to them. They obtain them only through the regularities that the distribution of primitive stuff throughout the whole of space and time exhibits.

To put the issue in terms of truth-makers, the distribution of primitive stuff throughout space and time makes true all the true propositions about the world, including in particular the propositions expressing laws of nature. Hence, if the laws of classical mechanics figure in the best system, predicates such as "mass" and "charge" apply to the particles in virtue of the patterns that the particle trajectories in space and time exhibit. These predicates – as well as all the other ones appearing in the propositions that are true about the world – really apply, and the propositions really are true, there is nothing fictitious about them. But what there is – and hence what makes them true – is nothing over and above the distribution of primitive stuff throughout space and time.

By the same token, when it comes to quantum physics, if one assumes that there is something in nature that determines, guides or pilots the temporal development of the primitive stuff, then one is committed to the wave-function being or representing a real entity in nature over and above the local matters of particular fact that carries out such a determination or guidance. But if one sides with the Humean in rejecting that assumption, then there is not only no need, but also no possibility either to maintain that the wave-function is or represents a real entity in nature, since doing so would destroy Humean supervenience. The only difference between classical and quantum mechanics in that respect is that in classical mechanics, the Humean can afford the luxury of taking variables such as mass and charge to refer to intrinsic properties that are instantiated in nature, provided that these properties are stripped off the functional role that they play in the laws of classical mechanics being essential to them. In quantum mechanics, by contrast, the Humean can no longer afford such a luxury with respect to the wave-function.

However, as far as physics is concerned, both variables such as mass and charge on the one hand and the wave-function on the other only enter the physics through the role they play in the dynamical laws that describe the temporal development of the distribution of matter in space. Therefore, as far as metaphysics is concerned, the Humean has in neither case a cogent reason to commit herself to such variables representing anything in nature over and above how primitive stuff is distributed in space-time. Quantum mechanics, by ruling out that the wave-function could be or represent a local quality, merely reveals an option that there was all the time, namely the option of an ontology that subscribes only to a commitment to primitive stuff – in short, physicalism without properties.

Generally speaking, a Humean supervenience basis being constituted by the distribution of primitive stuff throughout the whole of space-time (variation of occupied and unoccupied space-time points) is as unassailable as a metaphysical thesis can be: it is sufficient as a supervenience basis for all the possible empirical evidence for a physical theory, since all that evidence consists in how matter is distributed in space-time (cf. Bell 1987: 166). More precisely, a variation made up by some space-time points being occupied while others are empty is the most simple supervenience basis that is sufficient to ground all the possible empirical evidence. It arguably also is necessary as a supervenience basis for any empirical evidence, given that it is not at all clear how empirical evidence consisting in the distribution of matter in space-time could supervene on a basis that is not spatio-temporal. All the difference between the physical theories then boils down to a difference about the variables that have to enter into the best system, that is, the system that achieves the best balance between being simple and being informative in describing the distribution of the primitive stuff in the space-time of the actual world. By way of consequence, contrary to what the position known as epistemic structural realism claims (Worrall 1989), there is continuity in the ontology of physics, but change in the structure of the fundamental physical theories in the history of modern science; this continuity obtains because there is no reason to subscribe to the idea of an – inaccessible – intrinsic nature of the physical objects.

Apart from the metrical relations among space-time points and the occupation of some of these points by primitive stuff, the Humean is free to assign to all theoretical entities a place in the system achieving the best balance between simplicity and informativeness – starting with mass and charge in classical mechanics and continuing with the wave-function in quantum mechanics. That treatment can surely be extended to whatever theoretical entities physical theories going beyond quantum mechanics may introduce: the justification for admitting these entities precisely consists in the fact that they increase the balance between simplicity and informativeness in capturing the distribution of matter in space-time. Of course, space-time (the network of spatio-temporal relations among points) may itself be a dynamical entity, as assumed in the general theory of relativity, instead of constituting a passive background, as assumed in classical and in quantum mechanics. Humeanism can accommodate a dynamical space-time in the following manner: considering an initial configuration consisting in metrical relations between space-time points and some of these points being occupied by primitive stuff, the further development of the metrical relations depends on how the stuff is distributed in that initial configuration.

*4.     Conclusion*

This paper has sought to establish two conclusions – one for Humeanism and one for the metaphysics of quantum physics – as well as a general moral. As far as Humeanism is concerned, far from refuting Humeanism, quantum physics strengthens Humeanism in obliging the Humean to abandon the ornament of the local matters of particular fact being endowed with qualities in the sense of qualitative, intrinsic properties. They are just primitive stuff, in the sense of the fact that some points of space-time are occupied, whereas others are empty. Doing away with that ornament removes the stock objections against Humeanism from quidditism and humility. The debate on Humeanism can thus focus on the central issue of whether or not there are cogent reasons for a commitment to objective modality in the sense of, as far as physics is concerned, there being something in nature that drives, guides or enforces a certain temporal development of matter, by contrast to a certain such development simply happening to occur.

Furthermore, as far as the metaphysics of quantum physics is concerned, the physics – that is, the empirical evidence for quantum entanglement and quantum non-locality – by no means commits us to subscribe to either one of the following two consequences: to admit that non-supervenient relations or irreducible structures are instantiated in nature, or, if one shrinks from such a commitment, to endorse the view that the extremely high-dimensional configuration space of the universe is the realm of physical reality, with everything being local in that space. As far as the empirical evidence is concerned, one can simply stay with a primitive ontology of local matters of particular fact consisting in the distribution of primitive stuff in ordinary space. This conclusion holds not only for quantum mechanics, but also for any other physical theory, as long as measurement outcomes consist in "local beables". Again, the point at issue is whether or not there are cogent reasons for a commitment to there being something in nature that determines, guides or enforces the temporal development of matter. If so, one either has to go for non-locality in the sense of there being non-local entities such as non-supervenient relations or irreducible structures being instantiated in ordinary space-time, or, if one wishes to stay local, pay the price of regarding configuration space instead of four-dimensional space-time as the realm of physical reality.[1] But if the Humean is right, there is no need to go either of these ways.

The general moral of these considerations is this one: since the attack by Ladyman and Ross (2007: ch. 1) on mainstream analytic metaphysics, there is a tendency in metaphysics of science to oppose what is called scientific or naturalized metaphysics to analytic metaphysics. For instance, in a paper that has the intention to mark the contrast between scientific and speculative ontology, Humphreys (2013) scorns Humean supervenience in these terms:

> As an exercise in theorizing, Lewis's attitude would be unobjectionable were the position seriously put to an empirical test. But it has not in the sense that science long ago showed that Humean supervenience is factually false. The claim that the physical world has a form that fits the constraints of Humean supervenience is incompatible with well-known and well-confirmed theoretical knowledge about the non-separability of fermions. It was empirically established before Lewis's position was developed that entangled states in quantum mechanics exist and do not supervene on what would in classical cases be called the states of the components. This feature of our world is sufficiently well confirmed as to make Humean supervenience untenable.
> (Humphreys 2013: 56-7)

However, as the argument in this paper has made clear, it is an illusion to think that physics in itself can refute a metaphysical stance such as Humeanism. Of course, physics and metaphysics go together in the enquiry into the constitution of the world; but there is no neo-

---

[1]    I skip here the possibility to retain locality in ordinary space-time by finding a loophole in the proof of Bell's theorem or the subsequent experiments. See Maudlin (2011) for the standard monograph on quantum non-locality, setting out why there is no reasonable prospect to find such loopholes.

Quantum Humeanism                                                                 13positivist one-way road from physics to metaphysics that would enable one to read metaphysical conclusions directly off from physical theories and experiments.

## References

Albert, David Z. (1996): "Elementary quantum metaphysics". In: J. T. Cushing, A. Fine and S. Goldstein (eds.): *Bohmian mechanics and quantum theory: an appraisal*. Dordrecht: Kluwer. Pp. 277-284.

Albert, David Z. (2013): "Wave function realism". In: D. Z. Albert and A. Ney (eds.): *The wave function: essays in the metaphysics of quantum mechanics*. Oxford: Oxford University Press. Pp. 52-57.

Allori, Valia, Goldstein, Sheldon, Tumulka, Roderich and Zanghì, Nino (2008): "On the common structure of Bohmian mechanics and the Ghirardi-Rimini-Weber theory". *British Journal for the Philosophy of Science* 59, pp. 353-389.

Allori, Valia, Goldstein, Sheldon, Tumulka, Roderich and Zanghì, Nino (2013): "Predictions and primitive ontology in quantum foundations: a study of examples". *British Journal for the Philosophy of Science*, doi: 10.1093/bjps/axs1048

Aspect, Alain, Dalibard, Jean and Roger, Gérard (1982): "Experimental test of Bell's inequalities using time-varying analyzers". *Physical Review Letters* 49, pp. 1804-1807.

Beebee, Helen (2000): "The non-governing conception of laws of nature". *Philosophy and Phenomenological Research* 61, pp. 571-594.

Beebee, Helen and Mele, Alfred (2002): "Humean compatibilism". *Mind* 111, pp. 201-223.

Bell, John S. (1987): *Speakable and unspeakable in quantum mechanics*. Cambridge: Cambridge University Press.

Black, Robert (2000): "Against quidditism". *Australasian Journal of Philosophy* 78, pp. 87-104.

Bohm, David (1951): *Quantum theory*. Englewood Cliffs: Prentice-Hall.

Bohm, David (1952): "A suggested interpretation of the quantum theory in terms of 'hidden' variables". *Physical Review* 85, pp. 166-193.

Callender, Craig (unpublished): "Discussion: the redundancy argument against Bohm's theory". *Manuscript*. http://philosophyfaculty.ucsd.edu/faculty/ccallender/publications.shtml

Colbeck, Roger and Renner, Renato (2012): "Is a system's wave function in one-to-one correspondence with its elements of reality?" *Physical Review Letters* 108, 150402.

Darby, George (2012): "Relational holism and Humean supervenience". *British Journal for the Philosophy of Science* 63, pp. 773-788.

de Broglie, Louis (1928): "La nouvelle dynamique des quanta". *Electrons et photons. Rapports et discussions du cinquième Conseil de physique tenu à Bruxelles du 24 au 29 octobre 1927 sous les auspices de l'Institut international de physique Solvay*. Paris: Gauthier-Villars. Pp. 105-132. English translation in G. Bacciagaluppi and A. Valentini (2009): *Quantum theory at the crossroads. Reconsidering the 1927 Solvay conference*. Cambridge: Cambridge University Press. Pp. 341-371.

Dürr, Detlef, Goldstein, Sheldon and Zanghì, Nino (2013): *Quantum physics without quantum philosophy*. Berlin: Springer.

Einstein, Albert, Podolsky, Boris and Rosen, Nathan (1935): "Can quantum-mechanical description of physical reality be considered complete?" *Physical Review* 47, pp. 777-780.

Esfeld, Michael (2004): "Quantum entanglement and a metaphysics of relations". *Studies in History and Philosophy of Modern Physics* 35, pp. 601-617.

Esfeld, Michael (2009): "The modal nature of structures in ontic structural realism". *International Studies in the Philosophy of Science* 23, pp. 179-194.

Esfeld, Michael, Lazarovici, Dustin, Hubert, Mario and Dürr, Detlef (2013): "The ontology of Bohmian mechanics". Forthcoming in *British Journal for the Philosophy of Science*. DOI 10.1093/bjps/axt019

French, Steven (2014): *The structure of the world. Metaphysics and representation*. Oxford: Oxford University Press.

French, Steven and Ladyman, James (2003): "Remodelling structural realism: quantum physics and the metaphysics of structure". *Synthese* 136, pp. 31-56.